\documentclass[superscriptaddress,aps,pra,twocolumn,showpacs,floatfix]{revtex4-2}
\usepackage[utf8]{inputenc}
\usepackage{graphicx,amsmath,amsfonts,amssymb}
\usepackage{color}
\usepackage[colorlinks=true, allcolors={blue}]{hyperref}
\usepackage{graphicx}
\usepackage{epstopdf}
\usepackage{float}
\usepackage{placeins}
\usepackage{lipsum}

\usepackage{silence}
\newcommand{\orcid}[1]{\href{https://orcid.org/#1}{\includegraphics[width=7pt]{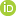}}}

\begin{document}

\preprint{APS/123-QED}

\title{The role of position momentum correlations in coherence freezing and purity behavior} 

%\title{Using position momentum correlations to control coherence freezing and purity}

%\title{Coherence and purity of mixed correlated Gaussian states}

\author{Pedro P. da Silva \orcid{000}}%\email{pedrop@ufpi.edu.br}
\affiliation{Departamento de F\'{i}sica, Universidade Federal do Piau\'{i}, Campus Ministro Petr\^{o}nio Portela, CEP 64049-550, Teresina, PI, Brazil.}

\author{Carlos H. S. Vieira \orcid{000}}\email{carloshsv09@gmail.com}
\affiliation{Centro de Ci\^{e}ncias Naturais e Humanas, Universidade Federal do ABC, Avenida dos Estados 5001, 09210-580, Santo Andr\'{e}, S\~{a}o Paulo, Brazil.}\affiliation{Departamento de F\'{i}sica, Universidade Federal do Piau\'{i}, Campus Ministro Petr\^{o}nio Portela, CEP 64049-550, Teresina, PI, Brazil.}

\author{Jonas F. G. Santos \orcid{000}}%\email{jonasfgs18@gmail.com}
\affiliation{Faculdade de Ci\^{e}ncias Exatas e Tecnologia, Universidade Federal da Grande Dourados,
Caixa Postal 364, Dourados, CEP 79804-970, MS, Brazil}

\author{Lucas S. Marinho \orcid{0000-0002-2923-587X}}\email{lucas.marinho@ufpi.edu.br}
\affiliation{Departamento de F\'{i}sica, Universidade Federal do Piau\'{i}, Campus Ministro Petr\^{o}nio Portela, CEP 64049-550, Teresina, PI, Brazil.}

\author{Marcos Sampaio \orcid{000}}%\email{marcos.sampaio@ufabc.edu.br}
\affiliation{Centro de Ci\^{e}ncias Naturais e Humanas, Universidade Federal do ABC, Avenida dos Estados 5001, 09210-580, Santo Andr\'{e}, S\~{a}o Paulo, Brazil}

\author{Irismar G. da Paz  \orcid{000}}\email{irismarpaz@ufpi.edu.br}
\affiliation{Departamento de F\'{i}sica, Universidade Federal do Piau\'{i}, Campus Ministro Petr\^{o}nio Portela, CEP 64049-550, Teresina, PI, Brazil.}

\date{\today}% It is always \today, today,
             %  but any date may be explicitly specified

\begin{abstract}
We explore the effects of Markovian bath coupling and initial position-momentum correlations on the coherence and purity of Gaussian quantum states. Our analysis focuses on the roles these factors play in the dynamics of quantum coherence, coherence lengths, and state purity. Our results reveal that initial position-momentum correlations have a remarkable impact on the quantum properties of the mixed state. These correlations lead to opposing behaviors in coherence and purity: as quantum coherence increases in response to stronger correlations, purity diminishes, and vice versa. This inverse relationship illustrates the phenomenon where, governed by these initial correlations, a state with greater mixing can display enhanced quantum coherence compared to a less mixed state. We also observe an unanticipated coherence freezing phenomenon, quantified by the relative entropy of coherence. Notably, this freezing is driven by initial position-momentum correlations, although the final frozen value is independent of these correlations.

\end{abstract}

\maketitle

\section{Introduction}\label{sec:intro}

%\tableofcontents
Quantum coherence is an essential ingredient for various forms of quantum correlations, including entanglement~\cite{Fan}, and serves as a vital physical asset in quantum computing~\cite{BoFan_PRL2023}, quantum information processing, and quantum thermodynamics~\cite{Jonas_PhysRevA2019,DeChiara_PhysRevResearch2024,Landi_PhysRevLett2019}. Understanding how distinct quantum features of a system (namely non-classical correlations, coherence, and purity) interrelate is crucial for both fundamental and experimental aspects \cite{Indrajith}. For instance,  the link between purity and maximum coherence was established through the use of unitary operations \cite{Streltsov}, showing that purity limits the highest achievable entanglement and discord, highlighting its fundamental role in quantum information processing. From the mathematical point of view, this study sheds light on two seemingly disconnected concepts: while quantum coherence is reflected in off-diagonal elements of a density matrix on a preferred basis, purity is measured based on the extent of deviations from the maximally mixed state besides being basis independent. 

Due to its basis-dependent nature, new approaches have been developed to quantify coherence in a manner invariant under arbitrary unitary transformations. For example, in Ref.~\cite{Byrnes}, a maximally mixed state (MMS) serves as the reference for an incoherent state. As the MMS is basis-invariant -- retaining its form across all bases -- a basis-independent measure of quantum coherence is established, linking directly to the concept of purity. Furthermore, this definition clearly distinguishes between collective and localized contributions in a way that remains consistent across different bases~\cite{Byrnes}.

On the other hand, two aspects are crucial in the study of coherence and purity as resources for quantum information science and are at the core of our study~\cite{Streltsov_2018,Luo2019,GOUR20151}.  First, the initial correlations present in the quantum state, involving parameters like position and momentum, along with its level of mixedness, can impact its quantum coherence. Second, and not less important, the interaction with the environment can lead to the loss of coherence, purity reduction, and information loss. Thus, it is crucial to effectively control and reduce the impact of interactions with the environment in quantum information processing tasks. One approach worth considering is the strategic manipulation of the quantum state's initial conditions to enhance its resilience against the influence of environmental factors. In \cite{Chen}, they study the impact of initial correlations and quantum coherence on the properties of a composite system by utilizing a bipartite two-level system that interacts through dipole-dipole interactions. It is shown that quantum coherence resulting from a relative phase in an initially correlated state exerts a substantial influence on energy transfer and information flow~\cite{Zicari_2023,Varizi_2021}. Also, the initial relative phase is seen to significantly affect steady-state entanglement (SSE) in the presence of an external laser field, allowing for the attainment of any SSE level with suitable phase choices.

Furthermore, the phenomenon of coherence freezing was observed, characterized by the complete invariance of coherence without any external control, arising under specific initial states and noisy evolutions~\cite{Plenio_AdessoRevModPhys2017, AdessoPRL2015,Adesso2015SciRep,Tong2016PRA,Adesso2016PRL}. This effect notably manifests in open quantum systems exposed to nondissipative Markovian decoherence channels~\cite{AdessoPRL2015}. It is crucial to distinguish the freezing of coherence from the concept of a decoherence-free subspace~\cite{LidarReview2014}, where open system dynamics effectively act as unitary evolution on a subset of states, thus preserving their informational properties. Experimentally, coherence freezing has been demonstrated even under Markovian bit-flip transversal noise, showcasing the resilience of quantum coherence~\cite{AdessoPRL2019}. Similarly, this effect was observed in the Gaussian Einstein–Podolsky–Rosen (EPR) entangled state transmitted through a Gaussian thermal noise channel~\cite{Kang2021Photon_Res}. Unlike in decoherence-free subspaces, in coherence freezing, the purity of the states may degrade while their coherence remains unaffected~\cite{Plenio_AdessoRevModPhys2017}. Experimentally, the robustness of quantum coherence of an optical cat state was demonstrated in the presence of losses, even as its negativity disappeared~\cite{Zhang2021Photonics}. Also, a method for quantum coherence freezing that directs quantum channels or the initial state (using a fixed-channel scheme) towards the coherence-freezing zone from any starting estimate was developed in \cite{GuoPRA2017}

In a previous study, we analyzed the time evolution of coherence quantifiers for a correlated Gaussian state uncoupled from its environment, highlighting a link between relative entropy of coherence and coherence lengths in position and momentum \cite{PP}. In this paper, we explore the dynamics of a mixed Gaussian state under system-environment coupling. We demonstrate that the diminution in coherence quantifiers and purity due to environmental effects is influenced by initial position-momentum correlations. Interestingly, these correlations inversely affect the coherence quantifiers and purity of the system studied. Additionally, our model exhibits a coherence freezing effect; while the purity of the mixed correlated Gaussian state degrades, its quantum coherence remains robust against strong environmental influences, aligning with observations in \cite{Plenio_AdessoRevModPhys2017}.

The organization of the work is the following. In Section \ref{sec:model}, we introduce the model for the system-environment interaction and calculate the master equation for the Markovian approximation. Then, we consider an initial Gaussian state produced by a real source, which carries source incoherent effects and position momentum correlations. We evolve the initial state with the propagator for the system-environment coupling and obtain the density matrix in position and momentum. In Section \ref{sec:coherence}, we calculate the elements of the covariance matrix, the coherence quantifiers, and the purity. 
In Section \ref{results}, we present our results for the wave dynamics of fullerene molecules through detailed plots. We analyze these plots in relation to the coupling with the environment and initial position-momentum correlations. Additionally, we derive an expression for quantum coherence in terms of purity, including its asymptotic behavior. This methodology allows us to estimate environment effects through reconstruction of the covariance matrix, from which we derive measures of quantum coherence and purity. Our final remarks and conclusions are presented in Section \ref{Conclusion}.

\section{Correlated Gaussian state of a particle interacting with a bath} \label{sec:model}
In this section, we describe the formalism of the master equation for a particle interacting with a Markovian bath and calculate the density matrix in position and momentum for an initial correlated Gaussian state. 

\subsection{System environment coupling and Master equation}

One possible way to describe a quantum system's behavior coupled with the environment is to apply the formalism of master equations \cite{Cohen_Tannoudji_atomphoton,Scully_book}. Consider the master equation for a central harmonic oscillator of frequency $\Omega$ interacting with an environment of harmonic oscillators. In the limit as $\Omega \to 0$, we encounter the scenario of a free particle in interaction with the environment, which constitutes the focus of our problem. The master equation, within the framework of the Born-Markov approximations, is provided by \cite{Schlosshauer2},
\begin{gather}
\dfrac{d}{dt}\hat{\rho}_{S}(t)=-i[\hat{H}^{\prime}_{S},\hat{\rho}_{S}]-\Lambda[\hat{x},[\hat{x},\hat{\rho}_{S}]]-
i\lambda[\hat{x},\{\hat{p},\hat{\rho}_{S}\}] \nonumber\\ -f[\hat{x},[\hat{p},\hat{\rho}_{S}]] \label{ME},
\end{gather}
where $\hat{H}^{\prime}_{S}=\hat{H}_{S}+(m\tilde{\Omega}^{2}\hat{x}^{2})/2$ is the frequency-shifted Hamiltonian of the system and $\hat{\rho}_{S}=\text{tr}_{E}(\hat{\rho}_{SE})$ describes the reduced density matrix of the system by tracing over the environmental degrees of freedom. Here we have defined the following coefficients~\cite{Schlosshauer2}:

\begin{gather}
\tilde{\Omega}^{2}=-\frac{2}{m}\int_{0}^{\infty}d\tau\eta(\tau)\cos(\Omega\tau),\;\Lambda=\int_{0}^{\infty}d\tau\nu(\tau)\cos(\Omega\tau), \nonumber\\ \lambda=\int_{0}^{\infty}\frac{d\tau\eta(\tau)\sin(\Omega\tau)}{m\Omega},
\;f=-\int_{0}^{\infty}\frac{d\tau\nu(\tau)\sin(\Omega\tau)}{m\Omega}. 
\end{gather}

To calculate the coefficients above, we can choose the linear relation for the spectral density of the bath oscillators with frequency $\omega$, i.e., $J(\omega)=\frac{2m\lambda_{0}}{\pi} \omega$ which is known as \textit{Ohmic spectral density} \cite{Schlosshauer2}. Here, $\lambda_{0}$ is the effective system-environment coupling. To prevent the spectral density from increasing indefinitely, a cutoff $\Delta$ is introduced, modifying the spectral density to $J(\omega) = \frac{2m\lambda_{0}}{\pi} \omega \frac{\Delta^{2}}{\Delta^{2} + \omega^{2}}$. It is evident that this cutoff effectively attenuates the spectral density for frequencies $\omega > \Delta$.

 Now, using this form of the spectral density, after  some mathematical manipulations we obtain the following results for the coefficients $\lambda$ and $\Lambda$:
\begin{equation}
\lambda=\frac{\Delta^{2}}{\Delta^{2}+\Omega^{2}}\lambda_{0},\;\;\Lambda=\frac{m\lambda_{0}\Omega\Delta^{2}}{\Delta^{2}+\Omega^{2}}\coth\left(\frac{\hbar\Omega}{2k_{B}T} \right). 
\end{equation}
For high temperature values, satisfying $\hbar\Omega \ll k_{B}T$, the approximation $\coth\left(\frac{\hbar\Omega}{2k_{B}T}\right) \approx \frac{2k_{B}T}{\hbar\Omega}$ holds true. Particularly in the case of a free particle where $\Omega \to 0$, this condition is readily met for any finite temperature value. Consequently, under the aforementioned limits and in the free particle scenario, we obtain $\lambda = \lambda_{0}$ and $\Lambda = \frac{2m\lambda_0 k_{B}T}{\hbar^{2}}$. Therefore, the parameter $\Lambda$ which depends on the bath temperature and the system environment coupling encode the environment effect on the quantum system. The coefficients $\tilde{\Omega}$ and $f$, which are simultaneously calculable, turn out to be negligible in these limits. Therefore, the temporal evolution of a system comprising a free particle coupled with an environment of oscillators is described by the master equation shown in equation (\ref{ME}). It is noteworthy that, depending on the time scale, the dissipative effect may be negligible, simplifying the master equation to
\begin{equation}
\dfrac{d}{dt}\hat{\rho}_{S}(t) = -i[\hat{H}_{S},\hat{\rho}_{S}] - \Lambda[\hat{x},[\hat{x},\hat{\rho}_{S}]].
\label{master_equation}
\end{equation}
The equation above produces the following propagator for the particle interacting with an environment made of gas at low pressure \cite{Joos}

\begin{gather}
G_{\Lambda}(x,x^{\prime},t;x_0,x^{\prime}_0,0)=\frac{m}{2\pi\hbar t}\;e^{\frac{im}{2\hbar t}\left[(x-x_0)^{2}+(x^{\prime}-x_0)^{2}\right]}\nonumber\\
\times e^{-\frac{\Lambda t}{3}\left[(x-x^{\prime})^{2}+(x_0-x^{\prime}_{0})^{2}+(x-x^{\prime})(x_0-x^{\prime}_0)\right]},
\label{progador}
\end{gather}
where $\Lambda t$ plays the role of the noise, such that the system becomes free from the environment if $\Lambda=0$ or $t=0$. Note that, the first exponential in Eq. (\ref{progador}) represents free dynamics, while the second takes into account the effects of the system's interaction with the environment, which will be responsible for the decoherence process~\cite{Joos,Viale2003PRA}.

We consider the propagation of fullerene molecules and the loss of coherence effect produced by air molecules scattering, as indicated in our theoretical model (see Fig. \ref{model_decohence}). In this case, the scattering constant is given by $\Lambda = (8/3\hbar^2)\sqrt{2\pi M}(k_B T)^{3/2}N w^2$ \cite{Schlosshauer2}. Here, $N$ is the total number density of the air, $M$ is the mass of the air molecule, $w$ is the size of the molecule of the quantum system, $k_B$ is the Boltzmann constant, and $T$ is the bath temperature. In deriving the master equation, Eq. (\ref{master_equation}), we assume that the center-of-mass state of the object remains undisturbed by the scattering event. This assumption is valid for fullerene molecules, which meet this condition. Furthermore, the scattering loss of coherence mechanism notably influences the fullerene's position, as evidenced by the second term in the master equation, $\Lambda[\hat{x},[\hat{x},\hat{\rho}_{S}]]$. This interaction also indirectly reduces the coherence length in momentum, which is detailed in Section \ref{results}.

\subsection{Correlated Gaussian state produced by a real source}

\begin{figure}[htp]
\centering
\includegraphics[width = 8.2 cm, height = 4.9 cm]{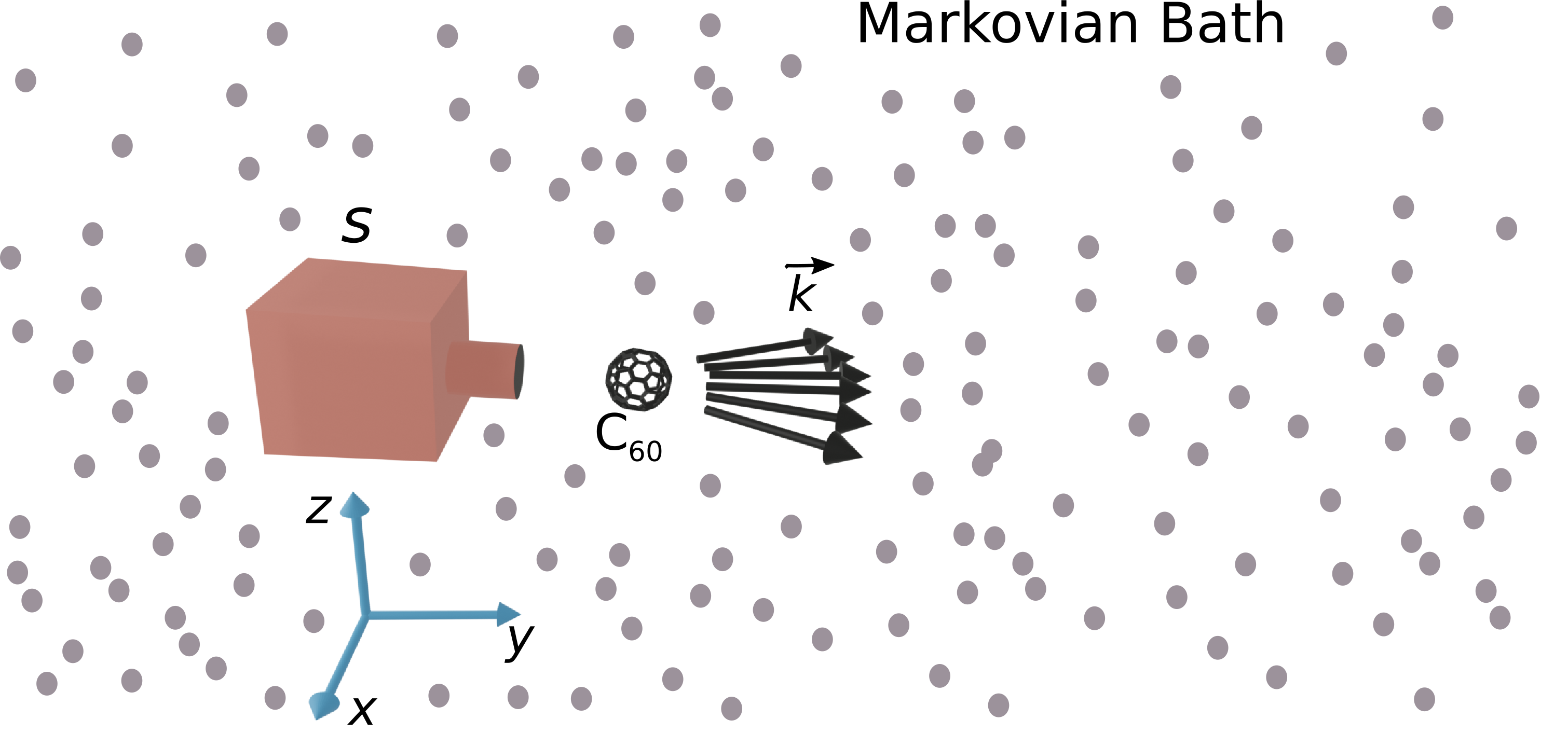}
\caption{Theoretical model. We consider a beam of fullerene molecules (C$_{60}$) ejected from an oven, resulting in a partially coherent source $S$ with coherence length encoded in the parameter $\ell_0$ and related with dispersion in momentum  $\vec{k}$ in the $Ox$ direction. The system is also interacting with a Markovian bath characterized by the environmental effect $\Lambda$. }
\label{model_decohence}
\end{figure}

The initial state under consideration is a correlated Gaussian state produced by a non-ideal source (as illustrated in Fig. \ref{model_decohence}) and described by \cite{Dodonov,Dodonov2,Yuen,PP}:
\begin{gather}
\rho_{0}(x_{0},x_{0}^{\prime}) = \frac{1}{\sqrt{\pi}\sigma_{0}} \exp\left[-(1-i\gamma)\frac{x_{0}^{2}}{2\sigma_{0}^{2}}\right] \nonumber\\
\exp\left[-(1+i\gamma)\frac{x_{0}^{\prime 2}}{2\sigma_{0}^{2}}\right] \exp\left[-\frac{(x_{0}-x_{0}^{\prime})^{2}}{2\ell_{0}^{2}}\right],
\label{i_s}
\end{gather}
where $\gamma$ denotes the position-momentum correlation parameter. The position-momentum correlation parameter $\gamma$ for the
the initial state is encoded as $\sigma_{xp}=(1/2)\langle xp+px\rangle-\langle x\rangle\langle p\rangle=\hbar\gamma/2$. Therefore, only when gamma is null we have an initial uncorrelated Gaussian state~\cite{Dodonov2}. The parameter $\ell_{0}$ is associated with the beam collimation in the transverse direction $Ox$, indicating the coherence level of the produced state. For the initial state (\ref{i_s}), complete coherence is achieved, i.e., $\ell_{0} \to \infty$, in the case of ideal collimation, whereas complete incoherence, i.e., $\ell_{0} \to 0$, results from the absence of collimation \cite{PP}.

Next, we apply the propagator for the system-environment coupling in the initial state (\ref{i_s}) to obtain the following density matrix in position representation, i.e., 
\begin{equation}
    \rho_{\Lambda}(x,x^{\prime},t)=\int \int dx_0 dx^{\prime}_0 G_{\Lambda}(x,x^{\prime},t;x_0,x^{\prime}_0,0)\rho_0(x_0,x^{\prime}_0).
    \label{rho_x}
\end{equation}
After some algebraic manipulations in Eq. (\ref{rho_x}) we obtain
\begin{equation}
\rho_{\Lambda}(x,x^{\prime},t)=\sqrt{\frac{2\bar{A}_{1}}{\pi}}\;e^{-\bar{A}_{1}(x^{2}+x^{\prime2})}\;e^{-\bar{A}_{2}(x-x^{\prime})^{2}}\;e^{i\bar{A}_{3}(x^{2}-x^{\prime2})},
\label{rho_xf}
\end{equation}
where $\bar{A}_1$, $\bar{A}_2$ and $\bar{A}_3$ are displayed in Appendix A. To obtain the density matrix in the momentum representation, a Fourier transformation is performed in Eq.~(\ref{rho_xf}), yielding
\begin{gather}
\rho_{\Lambda}(p,p^{\prime},t) =  \sqrt{\frac{2\bar{C}_{1}}{\pi}} \, e^{-\bar{C}_{1}(p^{2} + p^{\prime 2})} \, e^{-\bar{C}_{2}(p-p^{\prime})^{2}} \, e^{-i\bar{C}_{3}(p^{2} - p^{\prime 2})},
\label{MD_P}
\end{gather}
where the coefficients $\bar{C}_1$, $\bar{C}_2$, and $\bar{C}_3$ are also displayed in Appendix A.

\section{Loss of coherence and purity versus correlation}\label{sec:coherence}

Coherence is a fundamental resource in quantum physics, underlying many of its most puzzling and ubiquitous effects. Despite its significance, a framework for quantifying, characterizing, and manipulating coherence has only recently been proposed~\cite{Baumgratz2014}. This formalism has led to extensive exploration of coherence in various domains, including quantum thermodynamics \cite{Camati2019}, resource theory~\cite{Saxena2020}, and its connections to key quantum phenomena such as entanglement~\cite{Streltsov}, complementarity~\cite{Fan2019,Berrada}. In this section, we calculate some quantities that have been used to quantify the coherence of a quantum system and are related to the superposition principle, namely the relative entropy of coherence and the coherence length (CL) in position and momentum spaces. 

To quantify the coherence of a general quantum state $\rho$ (be it finite or infinite-dimensional), it is essential to define the reference basis. Baumgratz \textit{et al.} suggest that coherence can be quantified in terms of the off-diagonal elements using the $\ell_1$-norm, or via the relative entropy of coherence~\cite{Baumgratz2014}. Considering a state $\rho$, the relative entropy of coherence is defined as 
\begin{equation}
\mathcal{C}(\rho)=\underset{\delta\in\mathcal{I}}{\text{inf}}\left\{ S_{\text{KL}}(\rho||\delta)\right\},
\label{Rel_entr_coh}
\end{equation} 
where $S_{\text{KL}}(\rho||\delta)=\text{tr}(\rho\log_{2}\rho)-\text{tr}(\rho\log_{2}\delta)$ is the Kullback–Leibler (KL) divergence also called relative entropy and infimum runs over all incoherent states $\delta$~\cite{Baumgratz2014}. In the particular case in which the state is a $N$-mode Gaussian, Ref. \cite{Xu2016} established that the relative entropy of coherence (\ref{Rel_entr_coh}) can be written in terms of the first moments $\vec{d}$ and the covariance matrix $\boldsymbol{\sigma}$:
\begin{equation}
\mathcal{C}[\rho(\sigma,\vec{d})]=\mathcal{S}(\rho^{\text{th}})-\mathcal{S}(\rho),\label{eq:1}
\end{equation}
where the $\mathcal{S}(\rho)=-\text{tr}(\rho\ln\rho)$ is the von Neumann entropy, which for Gaussian states is set by 
\begin{equation}
\mathcal{S}(\rho)=-\sum_{j=1}^{N}\left[\frac{\nu_{j}-1}{2}\ln\left(\frac{\nu_{j}-1}{2}\right)-\frac{\nu_{j}+1}{2}\ln\left(\frac{\nu_{j}+1}{2}\right)\right],\label{eq:2}
\end{equation}
where $\left\{ \nu_{j}\right\} _{j=1}^{N}$ are the symplectic eigenvalues
of $\boldsymbol{\sigma}$ and $\rho^{\text{th}}$ is a $N$-mode reference thermal state
with average number of photons $\left\{ \bar{\varepsilon}_{j}\right\} _{j=1}^{N}$, $\vec{d} = (\langle q_1\rangle_\rho, ...,\langle p_1\rangle_\rho, \langle q_N\rangle_\rho, ...,\langle p_N\rangle_\rho)$, and $\sigma_{ij} = \langle x_{i} x_{j} + x_{j} x_{i}\rangle_\rho - 2\langle x_{i} \rangle_\rho \langle x_{j} \rangle_\rho$ are the elements of the covariance matrix. It is essential to highlight a key point concerning the quantification of coherence using the first and second moments of the state. Although the first and second moments are average values and thus independent of the basis choice, the selection of the reference state becomes crucial. We specifically choose the Fock basis (energy basis) to compute the reference state, which is typically a thermal state, as previously discussed.

In the case of a one-mode Gaussian state, the quantum coherence encoded in the Gaussian state $\rho_{\Lambda}$, calculated by the relative entropy of coherence (Eq.~\ref{eq:1}), reads~\cite{Xu2016}:
\begin{eqnarray}
\mathcal{C}(\rho_{\Lambda}) &=& \frac{\nu - 1}{2}\ln\left(\frac{\nu-1}{2}\right) - \frac{\nu + 1}{2}\ln\left(\frac{\nu+1}{2}\right)\nonumber\\ &+& (\bar{n}+1)\ln(\bar{n}+1) - \bar{n}\ln(\bar{n}),
\label{coherence_single_mode}
\end{eqnarray}
 where $\nu = 2\sqrt{\sigma_{11}\sigma_{22} - \sigma_{12}^2}$ is the determinant of the covariance matrix
\begin{equation}
\boldsymbol{\sigma}=\left(\begin{array}{cc}
\sigma_{11}  & \sigma_{12}\\
\sigma_{21} & \sigma_{22} \\
\end{array}\right),\label{eq:18}
\end{equation}
and $\bar{n} = (\sigma_{11} + \sigma_{22}+\vec{d}^{2} - 1)/2$ is the mean number of excitations of the single-mode system calculated using the Gaussian state $\rho_{\Lambda}$, with $\vec{d}=(\langle \hat{x}\rangle,\langle \hat{p}\rangle)$.

From the mixed Gaussian states seen in Eq. (\ref{rho_xf}) and Eq. (\ref{MD_P}), it is easy to check that the first moments $\langle\hat{x}\rangle$ and $\langle\hat{p}\rangle$ is null and the dimensionless second moments are given by
\begin{equation}
\sigma_{11}=\frac{t^{2}}{\tau_{0}^{2}}\left[\frac{1}{2}+2\left(\frac{\tau_{0}}{2t}+\frac{\gamma}{2}\right)^{2}+\frac{\sigma_{0}^{2}}{\ell_{0}^{2}}+\frac{2\Lambda t\sigma_{0}^{2}}{3}\right],
\end{equation}
\begin{equation}
\sigma_{22}=\frac{(1+\gamma^{2})}{2}+\frac{\sigma_{0}^{2}}{\ell_{0}^{2}}+2t\Lambda\sigma_{0}^{2}, 
\end{equation}
\begin{equation}
\sigma_{12}=\frac{\gamma}{2}+\frac{t}{2\tau_{0}}(1+\gamma^{2})+\frac{t\sigma_{0}^{2}}{\tau_{0}\ell_{0}^{2}}+\frac{t^{2}\Lambda\sigma_{0}^{2}}{\tau_{0}},
\end{equation}
with $\tau_{0}=m\sigma_{0}^{2}/\hbar$.

In the following, let us calculate the CL's which are also quantifiers of the coherence of a quantum system. The CL in position is defined as the width $x-x^{\prime}$ which involves off-diagonal elements of the density matrix. The square of the dimensionless CL is given by \cite{Joos}:
\begin{gather}
\ell_{x}^{2} (\gamma,\Lambda,t)=\frac{\langle \hat{x}^{2} \rangle}{\sigma_{0}^{2}} \mid_{\rho_{\Lambda}(x=-x^{\prime},t)} \nonumber\\
=\frac{2t^{2}\hbar^{2}}{\bar{D}}\Bigg[\frac{3}{2}+ 
\frac{\ell_{0}^{2}}{\sigma_{0}^{2}}\Big(t\Lambda\sigma_{0}^{2}+\frac{3}{4}(1+\gamma^{2})+\frac{3\tau_{0}}{4t}\big(2\gamma+\frac{\tau_{0}}{t}\big) \Big) \Bigg]
\label{coh_length_x}
\end{gather}

where
\begin{gather*}
\bar{D}=4\hbar^{2}\Lambda^{2}\ell_{0}^{2}\sigma_{0}^2t^4+12t\Lambda\Big[\ell_{0}^{2}m^{2}\sigma_{0}^{4}+t\hbar\sigma_{0}^{2}\left(m\ell_{0}^{2}\gamma+\frac{2t\hbar}{3}\right)  \nonumber\\
+\frac{1}{3}t^{2}\ell_{0}^{2}\hbar^{2}(\gamma^{2}+1)\Big] +3m^{2}\sigma_{0}^{2}(\ell_{0}^{2}+2\sigma_{0}^{2}).
\end{gather*}

% $$ %\bar{D}=4\hbar^{2}\Lambda^{2}\ell_{0}^{2}\sigma_{0}^2t^4+12t\Lambda\left[\ell_{0}^{2}m^{2}\sigma_{0}{}^{4}+t\hbar\sigma_{0}^{2}\left(m\ell_{0}^{2}\gamma+\frac{2t\hbar}{3}\right)+\frac{1}{3}t^{2}\ell_{0}^{2}\hbar^{2}(\gamma^{2}+1)\right]+3m^{2}\sigma_{0}^{2}(\ell_{0}^{2}+2\sigma_{0}^{2}).
% $$
Analogously, the momentum CL is defined as the width $p-p^{\prime}$ in the momentum space, and its (dimensionless) square is given by \cite{Joos}:
\begin{gather}
\ell_{p}^{2}(\gamma,\Lambda,t)=\frac{\sigma_{0}^{2}}{\hbar^{2}}\langle \hat{p}^{2} \rangle \mid_{\rho_{\Lambda}(p=-p^{\prime},t)} \nonumber\\
=\frac{3m^2\sigma_{0}^{4}}{2\bar{D}}\left[2+\frac{\ell_{0}^{2}}{\sigma_{0}^{2}}(1+\gamma^{2})+4t\Lambda\ell_{0}^{2}\right].
\label{coh_length_p}    
\end{gather}

Another interesting quantity to study here is the purity of the state,i.e., $\mu=\text{tr}\left(\rho^{2}\right)$. For our model, employing the Eq. (\ref{rho_xf}) or Eq. (\ref{MD_P}), we obtain the following result for this quantity
%\begin{widetext}
\begin{gather}
\mu^{-2}(\gamma,\Lambda,t)=1+\frac{4\sigma_{0}^{2}}{\ell_{0}^{2}}+4\sigma_{0}^{2}\Lambda t+\frac{4\gamma\Lambda\hbar}{m}t^{2}+ \nonumber \\
\frac{2\hbar\Lambda(3\gamma^{2}+2+\frac{4\sigma_{0}^{2}}{\ell_{0}^{2}})}{3\tau_{0}m}t^{3}+\frac{4\Lambda^{2}\hbar^{2}}{3m^{2}}t^{4}.
\label{purity}
\end{gather}
%\end{widetext}
This result produces $\mu=1$ when $\ell_0\rightarrow \infty$, i.e., for a completely coherent source and no environment effect $\Lambda=0$. Outside of this regime we always have $\mu(\gamma,\Lambda,t)<1$, typical of a mixed state. 

\section{Results and Discussion}\label{results}

\begin{figure*}[htb]
\centering
\includegraphics[scale=0.35]{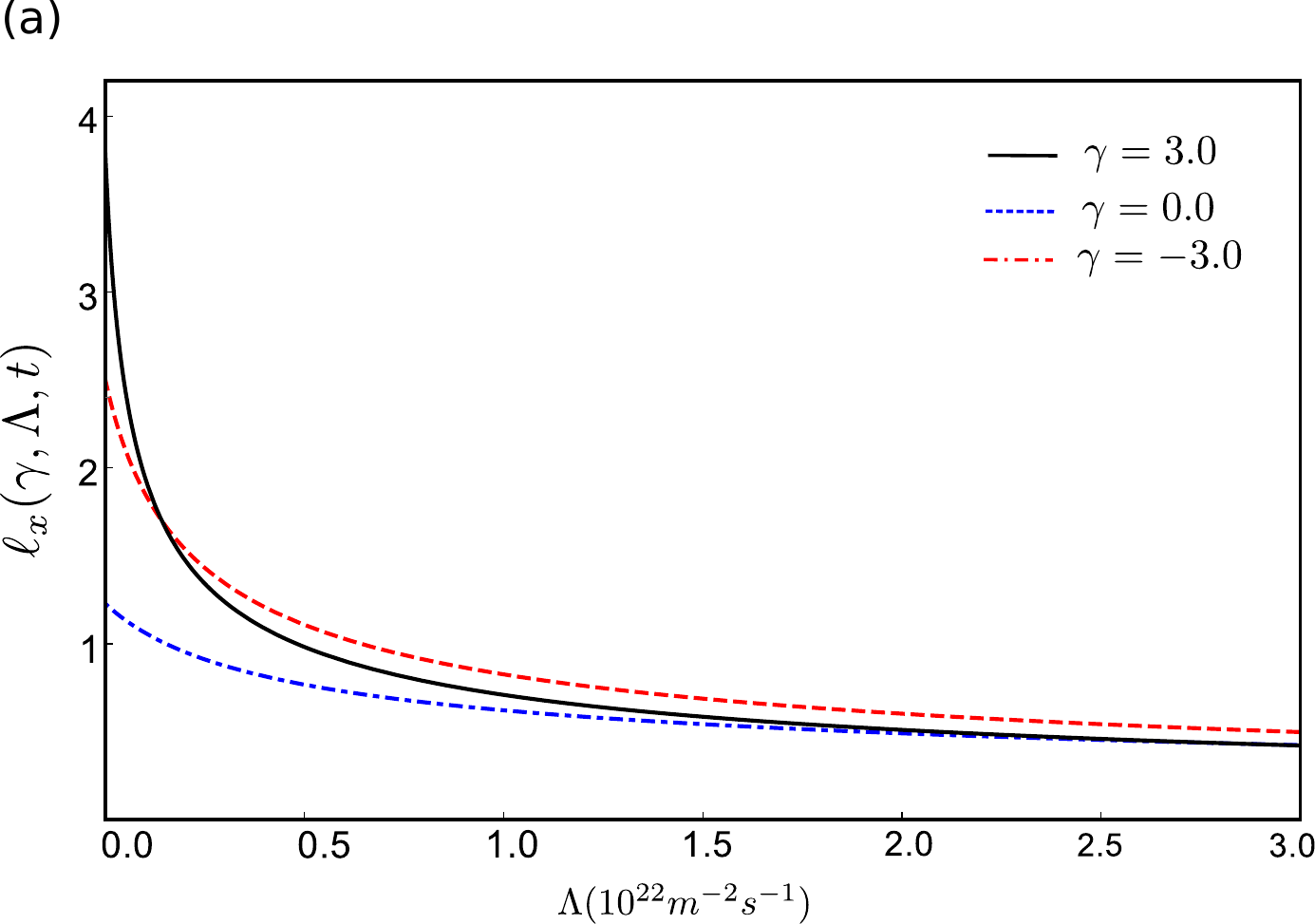}
\hspace{0.5cm}
\includegraphics[scale=0.35]{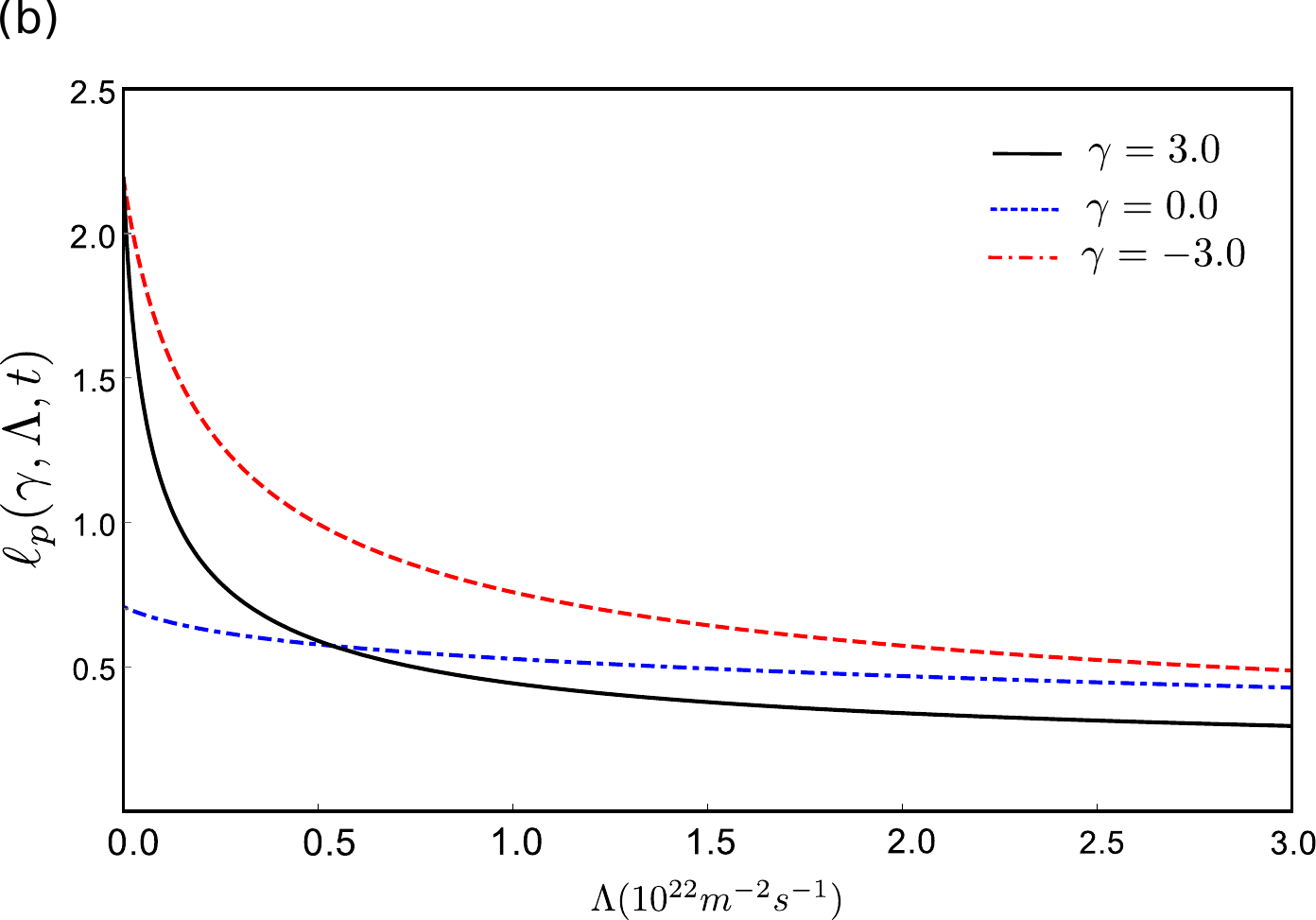}
\vspace{0.5cm}
\includegraphics[scale=0.35]{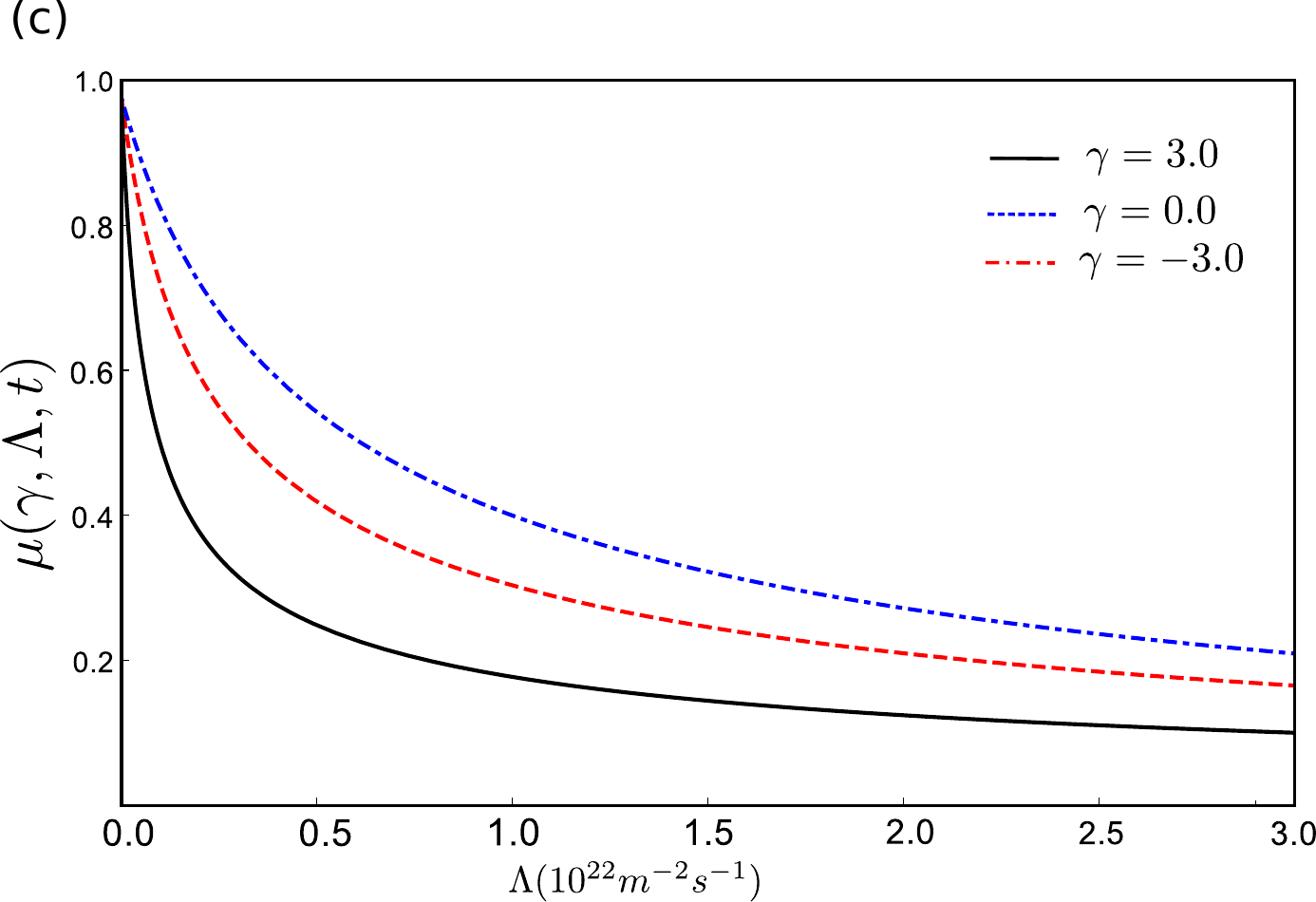}
\hspace{0.5cm}
\includegraphics[scale=0.35]{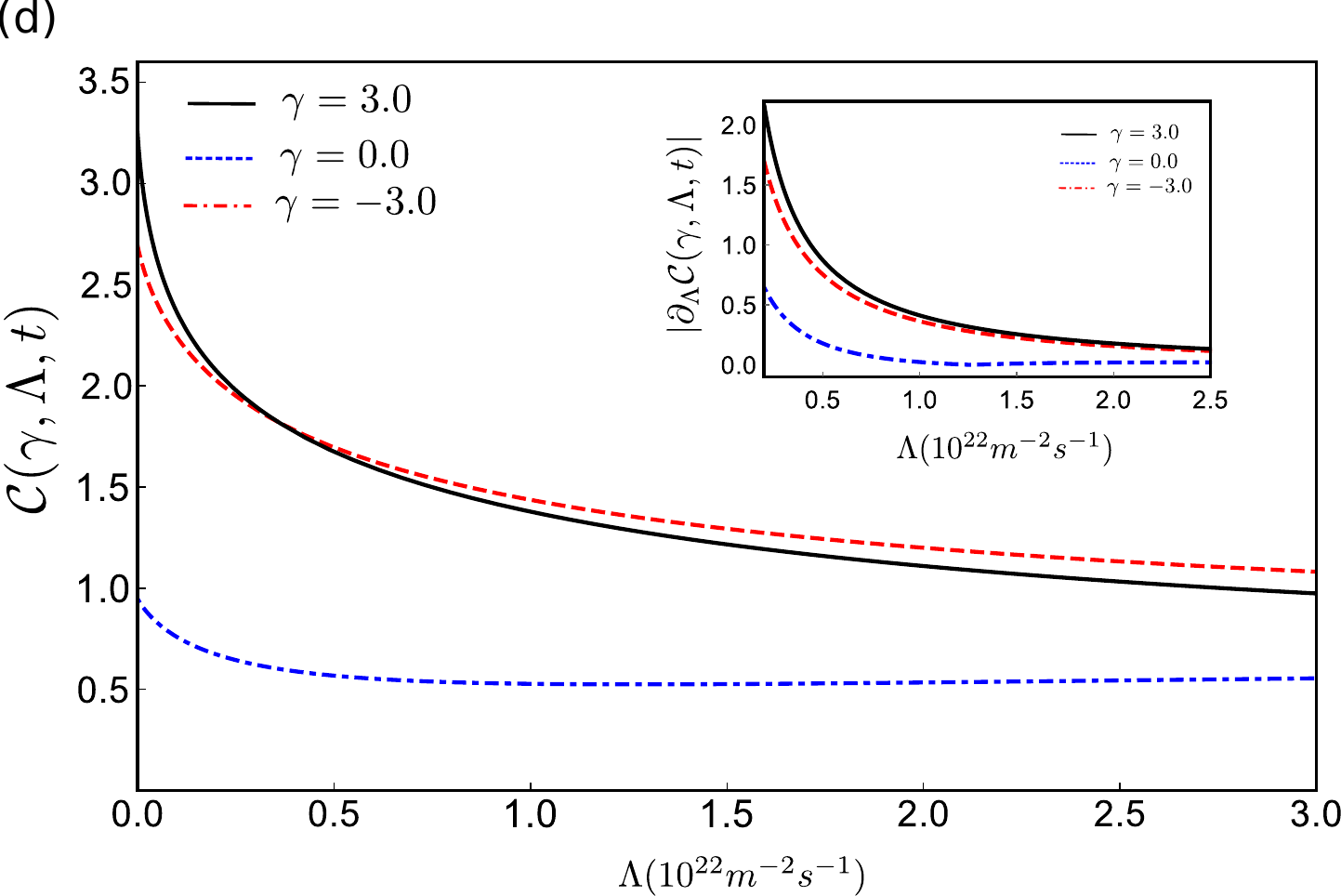}
\caption{Coherence quantifies as functions of environment effect $\Lambda$. (a) Coherence length in position normalized by $\sigma_{0}$, (b) coherence length in momentum normalized by $\hbar/\sigma_{0}$, (c) purity, and (d) quantum coherence for $t=1.0\;\mathrm{\mu s}$ and three values of $\gamma$.  We consider $\gamma=-3.0$ (red dashed line),  $\gamma=0.0$ (blue dotted line) and $\gamma=3.0$ (black solid line). The inset plot shows the absolute value of the rate of change of quantum coherence with environment effect $\Lambda$.}
\label{lx_dec}
\end{figure*}

We analyze the environment and initial position momentum correlations effects on fullerene molecules, adopting the
following parameters: fullerene mass $m=1.2 \times 10^{-24}$ kg,
molecular size $w=7$ $\mathring{A}$, width of the initial wave packet $\sigma_0 =7.8$ nm, mass of the air molecule $M=5.0 \times 10^{-26}$ kg, and the environment temperature $T = 300$ K \cite{Marinho_2018,Marinho_2020,Marinho_2023,Viale2003PRA}. It is assumed that the collimator apparatus selects wave numbers in the $Ox$ direction, with a transverse wavenumber dispersion of $\sigma_{k_{x}} \approx 10^{7}\;\mathrm{m^{-1}}$. This corresponds to an initial coherence length of $\ell_{0} = \sigma_{k_{x}}^{-1} \approx 50\;\mathrm{nm}$. Parameters of this order of magnitude were previously used in experiments with fullerene molecules by Zeilinger in Ref.~\cite{Zeilinger2002}.

Figure \ref{lx_dec} displays the quantifiers of coherence and purity as functions of the environment encoded by $\Lambda$, for a propagation time of $t=1.0\;\mathrm{\mu s}$ across three different values of the correlation parameter $\gamma$. Panel (a) shows the coherence length in position, (b) in momentum, (c) purity, and (d) quantum coherence. We analyze $\gamma=-3.0$ (red dashed line), $\gamma=0$ (blue dotted line), and $\gamma=3.0$ (black solid line).

The behavior of the coherence length in position as shown in figure \ref{lx_dec} (a) is explained by the terms in Eq. (18). At $\Lambda=0$, the denominator of Eq. (18) is independent of $\gamma$, while the numerator varies linearly and quadratically with $\gamma$. Consequently, $\gamma=3.0$ yields the highest $\ell_{x}$ value. Additionally, given $\tau_0/t<1$, $\gamma=-3.0$ results in a larger $\ell_{x}$ compared to $\gamma=0$. For $\Lambda\neq 0$, $\gamma=-3.0$ minimizes the denominator, thus maximizing the quantum coherence from a certain $\Lambda$ value onwards. The coherence length in momentum, $\ell_p$, shown in figure \ref{lx_dec} (b) follows from Eq. (19), where the numerator depends quadratically on $\gamma$ at $\Lambda=0$, making $\gamma=3.0$ and $\gamma=-3.0$ yield the same and highest values for $\ell_p$ compared to $\gamma=0$. With increasing $\Lambda$, $\gamma=-3.0$ continues to minimize the denominator, enhancing the quantum coherence. The purity, depicted in figure \ref{lx_dec} (c), is directly interpretable by examining the terms in Eq. (20). Finally, the quantum coherence behavior in figure \ref{lx_dec} (d) mirrors that of $\ell_{x}$,  which is a consequence of the fact that our model of loss of coherence is constructed for the position space.

The plots in figure \ref{lx_dec} demonstrate that all coherence quantifiers and the purity decrease with respect to the environment coupling $\Lambda$. Their asymptotic behaviors for large values of environment effect $\Lambda$ are given by
\begin{gather}
    \ell_x^2\rightarrow  \frac{1}{2t\Lambda}, \quad \ell_p^2\rightarrow \frac{3m^2}{2t^3\Lambda}, \quad \mu^2 \rightarrow \frac{3m^2}{4\hbar^2t^4\Lambda^2}, \nonumber\\
    \mathcal{C} \rightarrow \ln\left(\frac{\sqrt{3}\tau_{0}}{t}+\frac{\sqrt{3}}{3}\frac{t}{\tau_{0}}\right).
\end{gather}

While the coherence lengths and purity approach zero under strong effect of the environment, quantum coherence remains at a nonzero value. This effect, known as coherence freezing \cite{Plenio_AdessoRevModPhys2017}, demonstrates complete invariance of coherence despite increasing the effects of the environment, fulfilling the condition $\partial_{\Lambda} \mathcal{C}(\rho_{\Lambda}) = 0$. It is important to examine how initial position-momentum correlations influence the system's progression to this freezing regime, as depicted in the inset of figure~\ref{lx_dec}(d). For instance, nonzero $\gamma$ values lead to more resilient coherence, while for $\gamma = 0$, the freezing condition is reached with smaller environmental effects. Crucially, the ultimate freezing value is independent of these correlations.

\begin{figure*}[htp]
\centering
\includegraphics[scale=0.45]{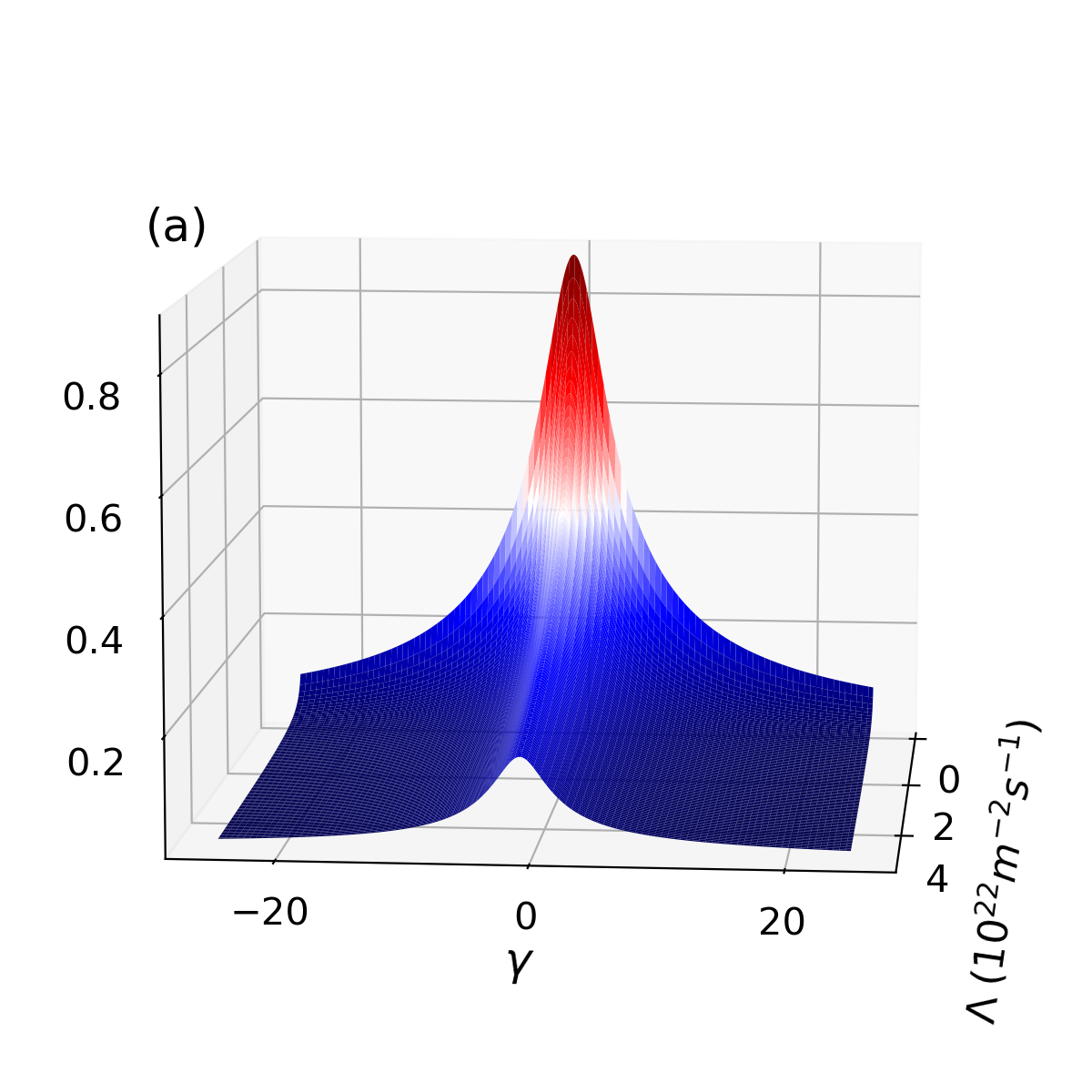} 
\includegraphics[scale=0.45]{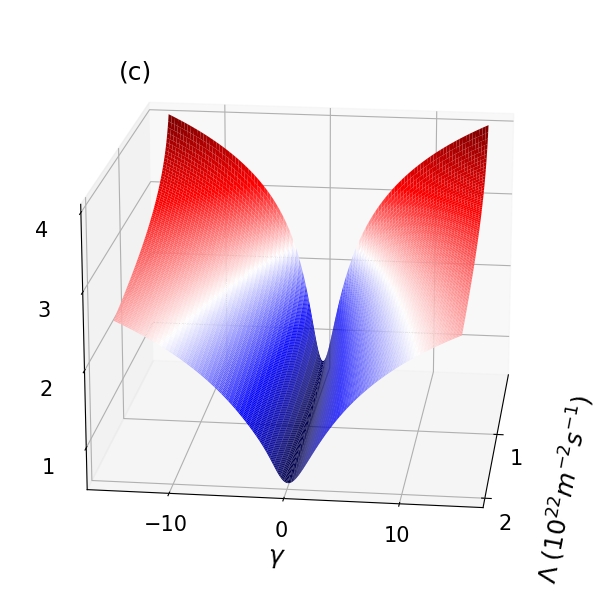}
\includegraphics[scale=0.45]{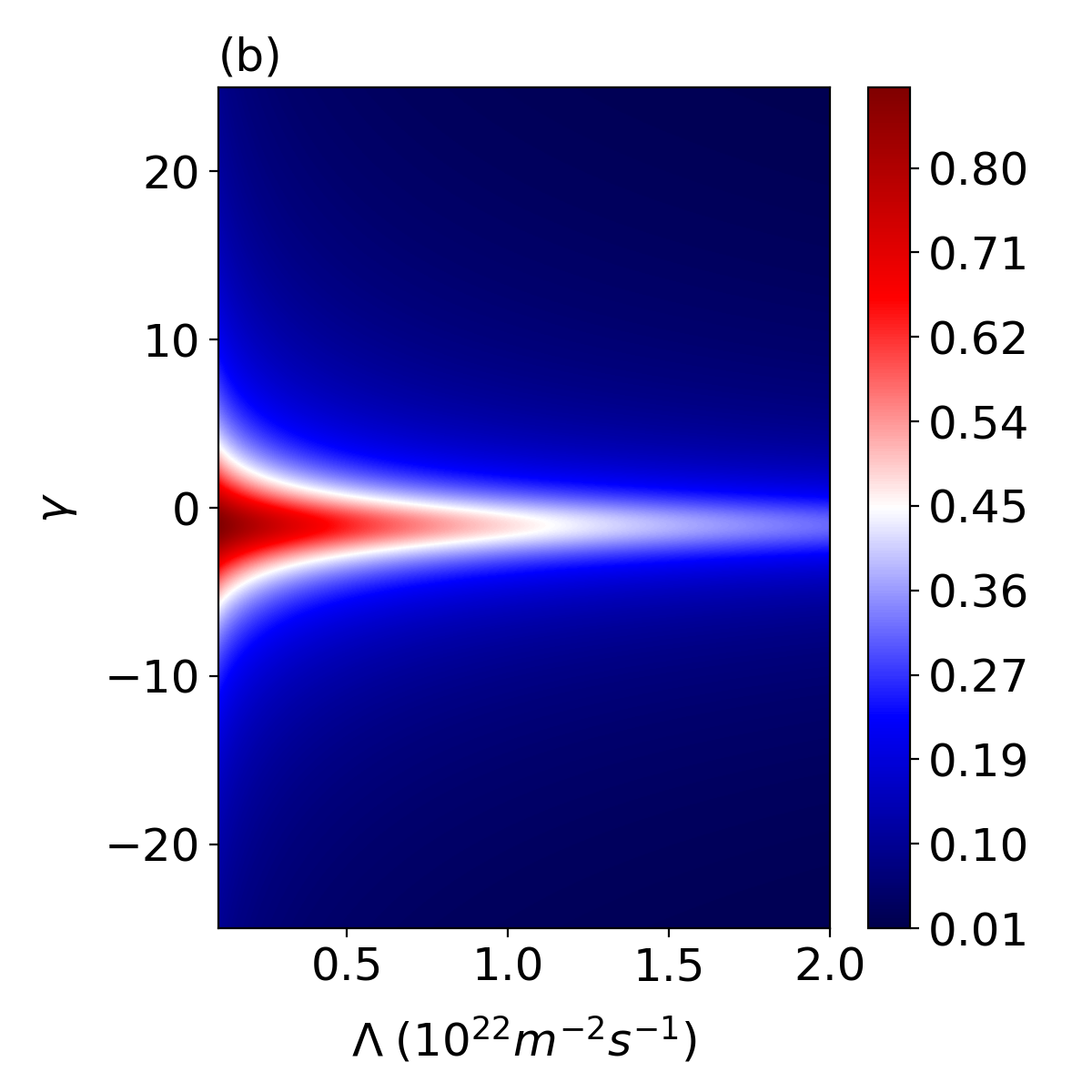}
\includegraphics[scale=0.45]{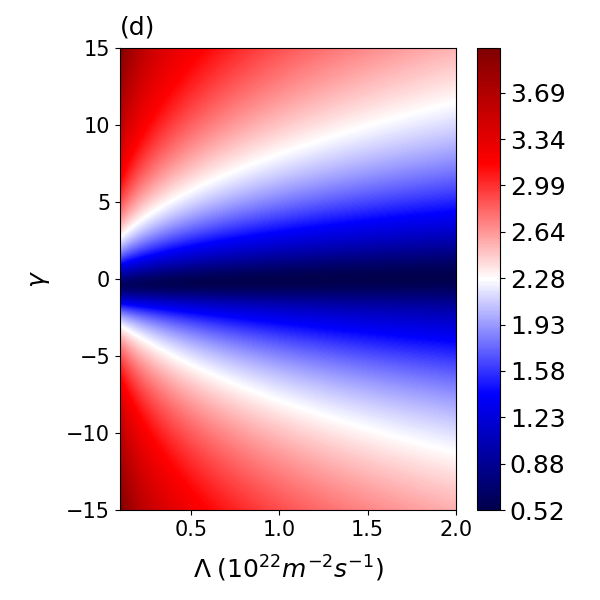}
\caption{Purity (left column) and the quantum coherence (right column) as a function of environment effect $\Lambda$ and correlation parameter $\gamma$ considering $t=1.0\;\mathrm{\mu s}$. (a) and (c) surface plots; (b) and (d) contour plots. }
\label{pureza_3d}
\end{figure*}

In figure~\ref{pureza_3d} (a) and (b), we present the surface and contour plots of the purity $\mu(\gamma,\Lambda,t)$, respectively, as functions of the environmental effect $\Lambda$ and the correlation parameter $\gamma$ for $t=1.0\;\mathrm{\mu s}$. In figure~\ref{pureza_3d} (c) and (d), we similarly display the surface and contour plots of quantum coherence, allowing for a comparison between the behaviors of purity and quantum coherence under the same conditions.
 
We note that purity decreases with increasing the effect of the environment and with large absolute values of the correlation parameter $\gamma$. Interestingly, purity exhibits a peak within a specific region of small absolute values of $\gamma$. Conversely, quantum coherence decreases with environmental effect but shows opposite behavior relative to purity concerning the correlation parameter. Specifically, quantum coherence reaches a minimum within the $\gamma$ interval where purity peaks and increases as the absolute values of $\gamma$ grow, while purity diminishes. The underlying reasons for these phenomena can be observed in the expressions for the elements of the covariance matrix and purity. In essence,
while large absolute values of the correlation parameter aid in incrementally enhancing
the quantum coherence through the elements of the covariance matrix, they concurrently intensify the environmental interaction in the expression of purity and contribute to its decrease.

\subsection{Relation between quantum coherence and purity}\label{section03}

\begin{figure*}[ht]
\centering
\includegraphics[scale=0.33]{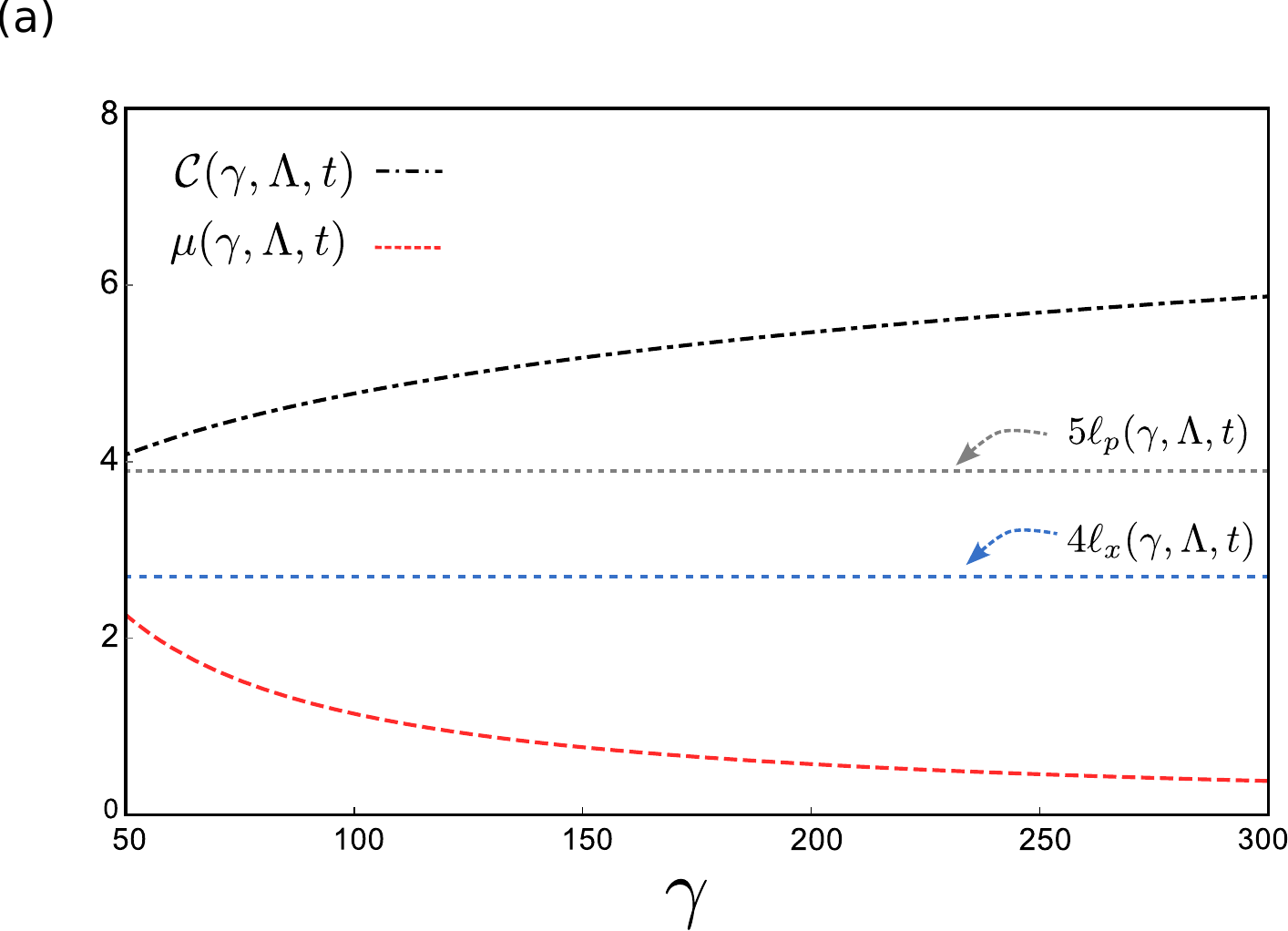}
\hspace{0.5cm}
\includegraphics[scale=0.33]{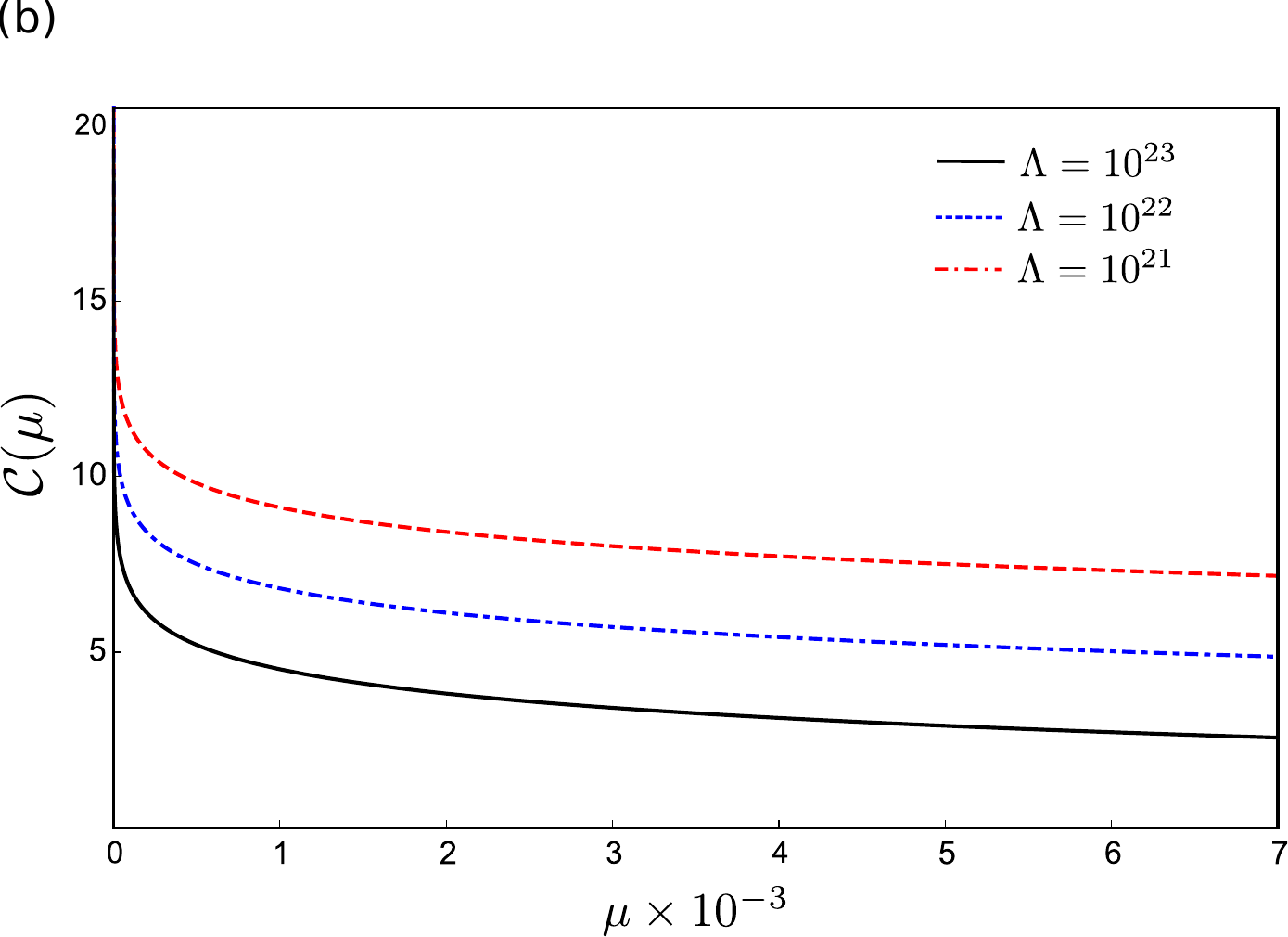}
\hspace{0.5cm}
\caption{(a) Quantum coherence, the coherence lengths, and purity as a function of the correlation parameter $\gamma$ for $\Lambda=10^{22}\;\mathrm{m^{-2}s^{-1}}$ and  $t=1.0\;\mathrm{\mu s}$. (b) Quantum coherence as a function of purity for $\gamma \rightarrow \infty$, $t=1.0\;\mathrm{\mu s}$, and three values of environment effect $\Lambda$. We consider  $\Lambda=10^{21}\;\mathrm{m^{-2}s^{-1}}$ (red dashed line),  $\Lambda=10^{22}\;\mathrm{m^{-2}s^{-1}}$ (blue dotted line), and $\Lambda=10^{23}\;\mathrm{m^{-2}s^{-1}}$ (black solid line).}
\label{coh_pureza}
\end{figure*}

We can show that quantum coherence can be expressed as a function of the coherence lengths and the purity. For the densities matrix (\ref{rho_xf}) and (\ref{MD_P}) we have $\langle \hat{x}\rangle=0$, $\langle \hat{p}\rangle=0$ and 

\begin{eqnarray}
\langle \hat{x}^{2}\rangle=\left(1+2\frac{\bar{A}_{2}}{\bar{A}_{1}}\right)\ell_{x}^{2}\;,\;\langle \hat{p}^{2}\rangle=\left(1+2\frac{\bar{C}_{2}}{\bar{C}_{1}}\right)\ell_{p}^{2}.
\label{eq22}
\end{eqnarray}
On the other hand, it is easy to show that the determinant of the covariance matrix is given by
\begin{eqnarray}
\nu^{2}=\left(1+2\frac{\bar{A}_{2}}{\bar{A}_{1}}\right)=\left(1+2\frac{\bar{C}_{2}}{\bar{C}_{1}}\right).
\label{eq23}
\end{eqnarray}
As it is known, the purity and the determinant of the covariance matrix are related by $\nu^{2}=\text{Det}(\sigma)=\mu^{-2}$ \cite{SerafiniPRA2003}. Thus, using the Eqs.(\ref{eq22}) and (\ref{eq23}) we can rewrite $\langle \hat{x}^{2}\rangle=\mu^{-2}\ell_{x}^{2}$ and $\langle \hat{p}^{2}\rangle=\mu^{-2}\ell_{p}^{2}$ which enable us to obtain $\bar{n}=\frac{1}{2\mu^{2}}(\ell_{x}^{2}+\ell_{p}^{2})-\frac{1}{2}$ and therefore the coherence Eq.~(\ref{eq:18}) becomes
\begin{widetext}
\begin{equation}
\mathcal{C}(\mu,\ell_{x},\ell_{p})=\frac{1}{2}\ln\left[\frac{4\mu^{2}(1-\mu)^{\frac{1-\mu}{\mu}}}{(1+\mu)^{\frac{1+\mu}{\mu}}}\right]+\frac{(\ell_{x}^{2}+\ell_{p}^{2})}{2\mu^{2}}\ln\left[\frac{(\ell_{x}^{2}+\ell_{p}^{2})+\mu^{2}}{(\ell_{x}^{2}+\ell_{p}^{2})-\mu^{2}}\right]+\frac{1}{2}\ln\left[\left(\frac{\ell_{x}^{2}+\ell_{p}^{2}}{2\mu^{2}}\right)^{2}-\frac{1}{4}\right].
\label{coh_purity}
\end{equation}
\end{widetext}

In figure \ref{coh_pureza} (a), we present quantum coherence, coherence lengths, and purity as functions of the correlation parameter $\gamma$, considering $\Lambda=10^{22}\;\mathrm{m^{-2}s^{-1}}$ and $t=1.0\;\mathrm{\mu s}$. It is observed that while quantum coherence increases, purity decreases. Coherence lengths also rise but stabilize at high values of $\gamma$. As Xu~\cite{Xu2016} has indicated, quantum coherence can be enhanced through coherent operations such as displacement and/or squeezing. Further, following the analysis by Marinho \textit{et al.} in \cite{Marinho_2020}, which examines the role of $\gamma$ in simulating a squeezing effect, we understand that the increase in quantum coherence with $\gamma$ can be attributed to this squeezing. This relationship is also evident in the increase of the elements of the covariance matrix $\sigma_{11}$, $\sigma_{22}$, and $\sigma_{12}$, reflecting the coherent contributions influenced by $\gamma$.

Conversely, the observed reduction in purity as a function of $\gamma$ can be elucidated by examining the fourth and fifth terms on the right-hand side of Eq. (\ref{purity}), where the parameter $\gamma$ appears multiplied by the environment effect parameter $\Lambda$. Consequently, an increment in $\gamma$ effectively amplifies the term $\gamma \Lambda$, which in turn amplify the mixed level of the system state, thereby diminishing the purity. 

In the limit as $\gamma \rightarrow \infty$, we derive $\ell^{2}_{x}=\frac{3}{8\sigma^{2}_{0}\Lambda t}$ and $\ell^{2}_{p}=\frac{3\tau_{0}^{2}}{8\sigma_{0}^{2}\Lambda t^{3}}$, values that are independent of $\gamma$. Consequently, quantum coherence, as expressed in Eq. (\ref{coh_purity}), can be formulated in terms of purity. It becomes evident that although quantum coherence and purity are both metrics for quantifying coherence properties, they exhibit distinct behaviors in relation to $\gamma$: the former increases while the latter decreases. In figure \ref{coh_pureza} (b), we illustrate quantum coherence from Eq. (\ref{coh_purity}) as a function of purity when $\gamma\rightarrow \infty$, for $t=1.0\;\mathrm{\mu s}$ and three environmental effects $\Lambda$. The values considered are $\Lambda=10^{21}\;\mathrm{m^{-2}s^{-1}}$ (red dashed line),  $\Lambda=10^{22}\;\mathrm{m^{-2}s^{-1}}$ (blue dotted line), and $\Lambda=10^{23}\;\mathrm{m^{-2}s^{-1}}$ (black solid line). At first glance, this might appear counterintuitive; however, figure \ref{coh_pureza} (b) presents a parametric plot between $\mathcal{C}$ and $\mu$ from figure \ref{coh_pureza} (a) and also demonstrates that an increase in system purity—achieved by reducing the initial position momentum correlation in this case—is not necessarily linked to an increase in quantum coherence \cite{Plenio_AdessoRevModPhys2017}.

 This seemingly paradoxical effect is also evident in the friendly qubit example, where quantum coherence in the energy basis is quantified by the $\ell_{1}$-norm as $\mathcal{C}_{\ell_1}(\rho_{\text{qubit}})=\sqrt{|\vec{r}|^2-z^2}$, and purity is defined as $\mu(\rho_{\text{qubit}})=|\vec{r}|$. Here, $|\vec{r}|$ represents the magnitude and $z$ the $z$-component of the Bloch vector. In the particular case where the state is initially totally mixed, represented in the Bloch sphere as $\vec{r}=(0,0,0)$, and subsequently evolves towards the north pole $(0,0,1)$ along the $z$-axis (path $\vec{r}=z\hat{z}$), the purity increases from zero to one while quantum coherence remains zero throughout this process. Conversely, if the same initial state evolves towards the equator of the Bloch sphere, both purity and quantum coherence increase, reaching a maximum value of one.

Additionally, for $\gamma >> 1$, the asymptotic behaviors of $\mathcal{C}(\gamma)$ and $\mu(\gamma)$ are, $\mathcal{C} = \ln (\gamma)$ and $\mu^2=\frac{\tau_{0}^{2}}{2\sigma_{0}^{2}t^3\Lambda\gamma^2}$. In this regime, purity can be parameterized by the correlation parameter, and quantum coherence can be expressed in terms of purity, with the inclusion of some additional constants, in a simplified form.

In this regime, purity can be parameterized by the correlation parameter, and quantum coherence can be expressed in terms of purity, modulo constant terms, in a simplified form:

\begin{figure}[ht]
\centering
\includegraphics[scale=0.37]{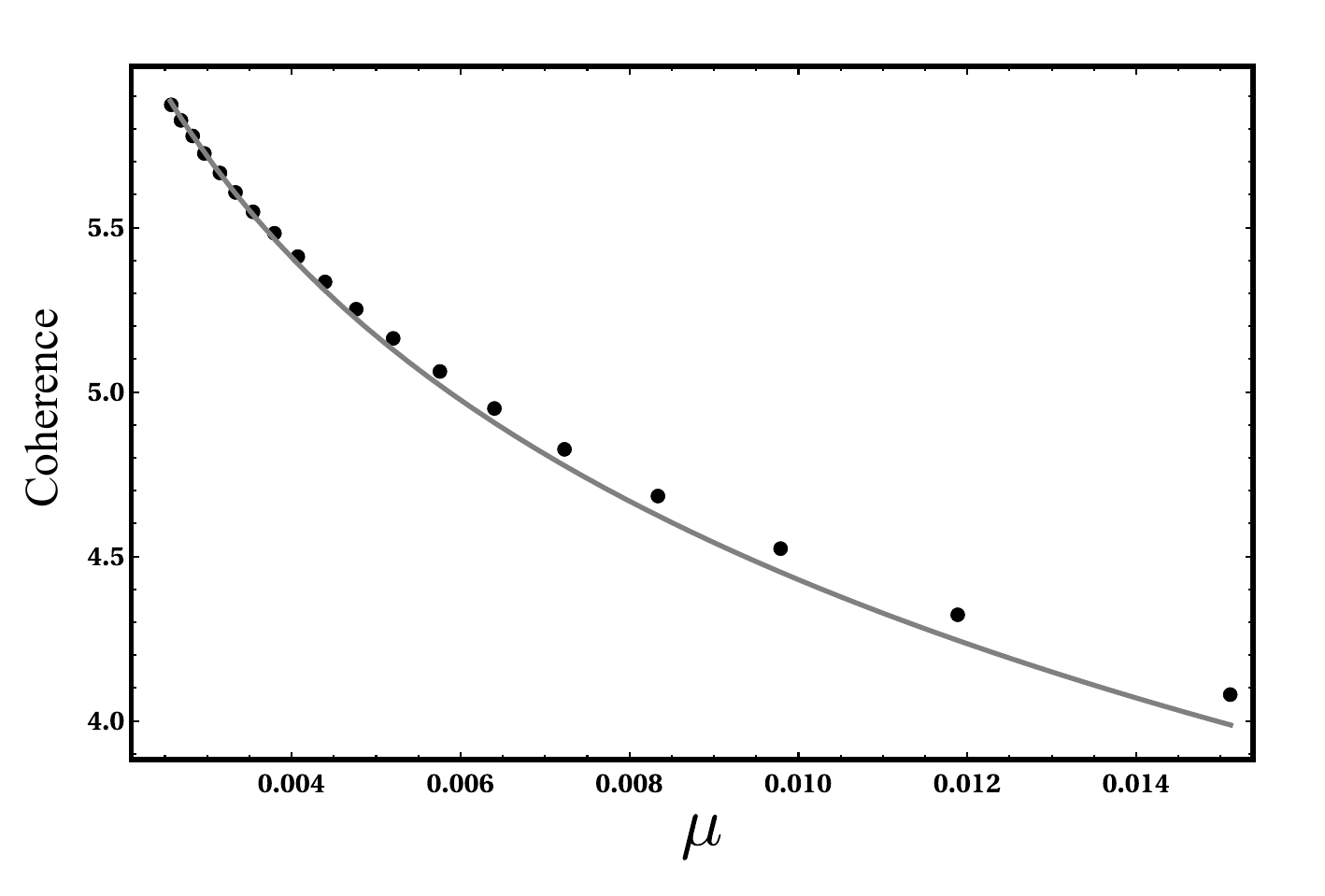}
\caption{Parametrization between the quantum coherence $\mathcal{C}$ and purity $\mu$ for large values of initial correlation parameter $\gamma$. The filled circles correspond to the full expression (\ref{coh_purity}) without any approximation. The solid line is an adjustment with the approximation (\ref{C_aprox}), where we include a small fitting parameter  $\varepsilon$  such that, $\mathcal{C}_\text{Fit}=\ln\left(\frac{\Theta(\Lambda)}{\mu}\right)^{1+\varepsilon}$, with $\varepsilon \approx 0.069$.}
\label{Fitting_figure}
\end{figure}

\begin{equation}\label{C_aprox}
\mathcal{C}(\mu)=\ln\left(\frac{\Theta(\Lambda)}{\mu}\right),\end{equation}
with $\mu <<1$ and  $\Theta(\Lambda)=\sqrt{\frac{\tau_{0}^{2}}{2\sigma_{0}^{2}t^{3}\Lambda}}$ being a constant that depends on the environment effect  $\Lambda$.  

In figure \ref{Fitting_figure}, we illustrate the relationship between quantum coherence $\mathcal{C}$ and purity $\mu$ for large initial correlation values $\gamma$. The filled circles represent the exact theoretical expression from Eq.(\ref{coh_purity}), while the solid line indicates a fit using the approximate Eq.(\ref{C_aprox}). To achieve this fit, we introduce a small fitting parameter $\varepsilon$, defined as $\mathcal{C}_\text{Fit}=\ln\left(\frac{\Theta(\Lambda)}{\mu}\right)^{1+\varepsilon}$, with $\varepsilon \approx 0.069$. This parameter reflects the slight deviation from the expected scaling relationship between $\mathcal{C}$ and $\mu$ due to our simplifications in the asymptotic behavior, though the deviation is minimal, i.e., $\varepsilon <<1$. The approximation is more accurate for lower values of purity $\mu$, which correspond to higher values of the correlation parameter $\gamma$, under which the parametrization was originally proposed. Table \ref{Tab} details the specifics of this parametrization. For example, when $\gamma = 105.6$, the approximation begins to lose validity, as demonstrated in figure \ref{Fitting_figure}.

\begin{table}[h]
\begin{centering}
\begin{tabular}{|c|c|c|}
\hline 
\textbf{Correlation $(\boldsymbol{\gamma})$} & \textbf{Purity $(\boldsymbol{\mu}\times 10^{-3})$} & \textbf{Coherence $(\boldsymbol{\mathcal{C}})$}\tabularnewline
\hline 
$50.0$ & $15.1$ & $4.1$\tabularnewline
\hline 
$105.6$ & 7.2 & $4.8$\tabularnewline
\hline 
$147.2$ & $5.2$ & $5.2$\tabularnewline
\hline 
$202.8$ & $3.8$ & $5.5$\tabularnewline
\hline 
$258.3$ & $3.0$ & $5.7$\tabularnewline
\hline 
$300.0$ & $2.6$ & $5.9$\tabularnewline
\hline 
\end{tabular}
\par\end{centering}
\caption{Parametrization of purity $\mu$ and quantum coherence $\mathcal{C}$ in terms of initial correlation parameter $\gamma$.}\label{Tab}
\end{table}

Furthermore, exploring the fit behavior of the quantum coherence $\mathcal{C}$ and the purity $\mu$, as depicted in FIG.~\ref{Fitting_figure}, is possible to assess the environment effect in the quantum system as follows
\begin{equation}
\Lambda(\mu,\mathcal{C})=\frac{1}{2t\sigma_{0}^{2}}\left[\frac{\tau_{0}}{ t\mu \exp{(\mathcal{C})}}\right]^{2}.
\label{Lamda(mu,C)}
\end{equation}

This expression is particularly valuable as it enables the experimental determination of properties of the environment (such as temperature and system-environment coupling) by analyzing the quantum coherence and purity of the mixed correlated Gaussian state. This analysis can be achieved through tomographic reconstruction of the covariance matrix. Specifically, in the case of fullerene molecules moving through an air molecule environment, information about the environment and the system-environment interaction can be deduced from the quantum coherence and purity of the fullerene molecules' mixed quantum state post-scattering by air molecules, by solely reconstructing the covariance matrix.

\section{Conclusion}\label{Conclusion}

We investigated coherence quantifiers and purity for a quantum state of fullerene molecules interacting with a Markovian bath. The initial state of the system is a correlated Gaussian state produced by a real source. Specifically, both purity and the coherence lengths asymptotically approach zero as a function of the effect of the environment, whereas quantum coherence remains constant in the limit of strong environmental impact. This phenomenon, known as the freezing of coherence, has been extensively studied in the literature. The rate at which this regime is reached is influenced by initial position-momentum correlations (initial conditions), although the frozen value remains independent of these correlations.

We observed that purity decreases with the effect of the environment but reaches maximum values in regions of small absolute values of initial position-momentum correlations. Conversely, purity decreases for large absolute values of the initial correlations. Quantum coherence also decreases with the effect of the environment but exhibits behavior opposite to that of purity with respect to the initial correlations. In regions where purity increases, quantum coherence decreases, and vice versa. As outlined in our theoretical model, the scattering-induced loss of coherence mechanism directly influences the tracking of fullerene's position. Strong position-momentum correlations, indicating a correlation between large momentum values and significant position values, diminish the monitoring of the particle's position, thereby reducing the loss of coherence. Conversely, correlations with large momentum values increase the mixture level of the quantum state, thus reducing its purity. For large values of initial position-momentum correlations, we derived a simple relationship between quantum coherence and purity that allows for inference of environmental properties and system-environment coupling by reconstructing the covariance matrix of the mixed state of fullerene molecules after scattering by air molecules.

Finally, we wish to emphasize the role of single-particle position-momentum correlations (defined as covariance $\sigma_{xp}$) in the quantum quantities studied in this problem. In the context of a composite system, such as a bipartite system, these correlations are akin to entanglement correlations. However, these quantum correlations for a single particle have not yet been related to entanglement or discord and have not been deeply explored in quantum systems. Here, we demonstrated that these correlations are crucial in shielding quantum coherence from the effect of the environment. Depending on the interval, these correlations can protect either purity or quantum coherence from the effect of the environment, but not both quantities simultaneously. In essence, the presence of quantum coherence can indicate the existence of single-particle position-momentum quantum correlations.

\section*{Acknowledgments}
P.P.S Thanks PROPESQ (PPGF/UFPI/PI). C.H.S.V. acknowledges CAPES (Brazil) and the Federal University of ABC for the financial support. J. F. G. Santos acknowledges CNPq Grant No. 420549/2023-4, Fundect and Universidade Federal da Grande Dourados for support. L.S.M. acknowledges the Federal University of Piau\' i for providing the workspace. M.S. acknowledges a research grant 302790/2020-9 from CNPq.  I.G.P. acknowledges Grant No. 306528/2023-1 from CNPq. 

\section*{Data Availability}
All data generated or analyzed during this study are included in this published article (and its supplementary information files).

\section*{Appendix A: Parameters of the density matrices and second moments}
The parameters of the density matrix in position and momentum are given by:
\begin{equation}
\bar{A}_{1}=\frac{m^{2}}{8\hbar^{2}t^{2}\sigma_{0}^{2}\bar{B}^{2}},
\end{equation}

\begin{eqnarray}
\bar{A}_{2}&=&\frac{m^{2}}{4\hbar^{2}t^{2}\bar{B}^{2}}\left(\frac{1}{2l_{0}^{2}}+\Lambda t\right)+\frac{\Lambda t}{12\sigma_{0}^{2}\bar{B}^{2}}\left(\frac{1}{\sigma_{0}^{2}}+\frac{2}{l_{0}^{2}}+\Lambda t\right)\nonumber\\
&+&\frac{m\Lambda \gamma}{4\hbar \sigma_{0}^{2}\bar{B}^{2}}+\frac{\Lambda t\gamma^{2}}{12\sigma_{0}^{4}\bar{B}^{2}},
\end{eqnarray}
\begin{eqnarray}
\bar{A}_{3}&=&\frac{m}{4 \hbar t\sigma_{0}^{2}\bar{B}^{2}}\left(\frac{1}{2\sigma_{0}^{2}}+\frac{1}{l_{0}^{2}}+\Lambda t\right)\nonumber\\
&+&\frac{m\gamma}{8\hbar t \sigma_{0}^{2}\bar{B}^{2}}\left(\frac{m}{\hbar t}+\frac{\gamma}{\sigma_{0}^{2}}\right),
\end{eqnarray}
and

\begin{eqnarray}
\bar{B}^{2}=\frac{1}{4\sigma_{0}^{4}}+\frac{1}{2\sigma_{0}^{2}l_{0}^{2}}+\frac{\Lambda t}{3\sigma_{0}^{2}}+\left(\frac{m}{2\hbar t}+\frac{\gamma}{2\sigma_{0}^{2}}\right)^{2}.
\end{eqnarray}

\begin{eqnarray}
\bar{C}_1=\frac{\bar{A}_1}{4\hbar^{2}(\bar{A}_{1}^{2}+2\bar{A}_1\bar{A}_2+\bar{A}_{3}^{2})},\bar{C}_2=\frac{\bar{A}_2}{\bar{A}_{1}}\bar{C}_1,\bar{C}_3=\frac{\bar{A}_3}{\bar{A}_{1}}\bar{C}_1\nonumber
\end{eqnarray}

The dimensionless second moments are defined by
\begin{eqnarray}
\sigma_{11}&=&\frac{\langle \hat{x}^{2}\rangle}{\sigma_{0}^{2}}=\frac{1}{\sigma_{0}^{2}}Tr[\rho_{\Lambda}(x,x^{\prime},t)\hat{x}^{2}],
\end{eqnarray}

\begin{eqnarray}
\sigma_{22}&=&\frac{\sigma_{0}^{2}}{\hbar^{2}} \langle \hat{p}^{2}\rangle=\frac{\sigma_{0}^{2}}{\hbar^{2}}Tr[\rho_{\Lambda}(p,p^{\prime},t)\hat{p}^{2}],
\end{eqnarray}
and 
\begin{eqnarray}
\sigma_{12}&=&\frac{1}{2\hbar} \langle (\hat{x}\hat{p}+\hat{p}\hat{x})\rangle=\frac{1}{2\hbar}Tr[\rho_{\Lambda}(x,x^{\prime},t)(\hat{x}\hat{p}+\hat{p}\hat{x})]. \nonumber \\
\end{eqnarray}

\bibliography{references}% Produces the bibliography via BibTeX.

\end{document}